\documentclass[12pt]{article}

\usepackage{a4}
\usepackage{latexsym,amssymb}
\usepackage{citesort}

\newcommand{\hso}{hypersurface--orthogonality }

\newcommand{\mystar}{*}
\newcommand{\Uomega}{{\cal U}}
\newcommand{\hE}{\hat E}
\newcommand{\hh}{\hat h}
\newcommand{\hR}{\hat R}
\newcommand{\hm}{\hat m}
\newcommand{\hQ}{\hat Q}
\newcommand{\hSigma}{\hat \Sigma}
\newcommand{\beq}{\begin{equation}}
\newcommand{\eeq}{\end{equation}}
\newcommand{\beqa}{\begin{eqnarray}}
\newcommand{\eeqa}{\end{eqnarray}}
\newcommand{\cref}[1]{4\emph{\ref{#1})}}
\newcommand{\Dh}{\Delta_{h}}
\newcommand{\RN}{Reissner--Nordstr\"om\ }
\newcommand{\sMP}{standard Majumdar--Papapetrou\ }
\newcommand{\MP}{Majumdar--Papapetrou\ }
\newcommand{\hS}{\hat{\Sigma}}

\newcommand{\cU}{{\cal U}}
\newcommand{\cUc}{{\cal V}}

\newcommand{\R}{\Bbb R}
\newcommand{\ext}{{\mathrm{ext}}}
\newcommand{\Sext}{\Sigma_\ext}
\newcommand{\Snd}{\partial_{nd}\Sigma}

\newcommand{\proof}{\noindent{\sc Proof:}\ }

\newtheorem{Theorem}{Theorem}[section]
\newtheorem{Corollary}[Theorem]{Corollary}
\newtheorem{Lemma}[Theorem]{Lemma}
\newtheorem{Proposition}[Theorem]{Proposition}

\newcommand{\QED}
   {\hfill$\hbox{\vrule height1.3ex width1.3ex depth.1ex}\ $
     \\ \medskip}
\newcommand{\qed}{\QED}
\newcommand{\eq}[1]{(\ref{#1})}

 \def\scri{\hbox{${\cal J}$\kern -.645em {\raise
      .57ex\hbox{$\scriptscriptstyle (\ $}}}}

\newcommand{\ks}{{\cal K}(\Sigma)}

\newcommand{\Mext}{M_{\mathrm{ext}}}
\newcommand{\doc}{{\cal D}_{oc}}
\newcommand{\pdoc}{\partial\doc}

\newcommand{\be}{\begin{equation}}
\newcommand{\ee}{\end{equation}}

{\catcode `\@=11 \global\let\AddToReset=\@addtoreset}
\AddToReset{equation}{section}

\newcounter{mnotecount}[section]

\newcommand{\oldmnote}[1]{ \marginpar{\raggedright\tiny\em old mnote
  in the file here  (to be discarded as far as PC is   concerned)}}
\newcommand{\oldnote}[1]{}  
\newcounter{pcheckcount}[section]

\newcommand{\pcheck}[1]{}

\begin{document}

\title{Towards a classification of static  electro--vacuum
  space--times containing an asymptotically flat spacelike
  hypersurface with compact interior}

\author{
Piotr T.\ Chru\'sciel\protect\thanks{%
  Alexander von Humboldt fellow.  Supported in part by a grant from
  the Polish Committee for Scientific Research (KBN) No 2 PO3B 073 15
  and by the Humboldt Foundation.
  \emph{Email}: Chrusciel@Univ-Tours.Fr} \\
D\'epartement de Math\'ematiques\\
Facult\'e des Sciences\\
Parc de Grandmont\\
F37200 Tours, France\\
}\maketitle

\begin{abstract}
  We show that static electro--vacuum black hole space--times
  containing an asymptotically flat spacelike hypersurface with
  compact interior and with both degenerate and non--degenerate
  components of the event horizon do not exist, under the
  supplementary hypothesis that all degenerate components of the event
  horizon have charges of the same sign. This extends previous
  uniqueness theorems of Simon~\cite{Simon:elvac} and Masood--ul--Alam~\cite{Masood} (where only non--degenerate horizons were allowed)
and
  Heusler~\cite{heuslerMP} (where only degenerate horizons were
  allowed).
\end{abstract}

\vfill\eject
\section{Introduction}
\label{SI}

A classical question in general relativity, first raised and
partially answered by Israel~\cite{Israel:elvac}, is that of
classification of suitably regular static\footnote{\label{statfoot}A
space--time $(M,g)$
  with Killing vector field $X$ will be called static if $X$ is
  (locally) hypersurface orthogonal everywhere, and if $X$ is timelike
  for all sufficiently distant points in the relevant asymptotically
  flat regions, \emph{cf.} Section 2. The regions of $M$ where $X$ is
  timelike are thus static in the usual sense. It should be emphasized
  that we allow the defining Killing vector of a static space--time to
  be spacelike in some regions.  Thus both the Schwarzschild
  space--time and its Kruszkal--Szekeres extension are static in our
  terminology. We hope that this will not lead to confusions.} black
hole solutions of the Einstein--Maxwell equations.  The most
complete results existing in the literature so far are due to Simon
\cite{Simon:elvac}, Masood--ul--Alam~\cite{Masood} and Heusler
\cite{HeuslerRNuniqueness,heuslerMP} who show, roughly speaking, the
following:
\begin{enumerate}
\item Suppose that all the horizons are non--degenerate. Then the
  black hole is a Reissner--Norsdstr\"om black hole
~\cite{Simon:elvac,Masood,HeuslerRNuniqueness}.
\item Suppose instead that all the horizons are degenerate, and that
  \begin{equation}
    \label{eq:qc}
\forall\ i,j\qquad    Q_i Q_j \ge 0\ ,
  \end{equation}
  where $Q_i$ is the charge of the $i$--th connected component of the
  black hole.  Then the black hole is a \sMP
  black hole~\cite{heuslerMP} (\emph{cf.\/} also
~\cite{ChAsconaa,ChNad}).
\end{enumerate}
Heusler's condition \eq{eq:qc} is obviously satisfied by a connected
black hole, so that the above results settle the classification
question in the connected case (recall that a standard connected
Majumdar--Papapetrou black hole is an extreme Reissner--Nordstr\"om
one). The general case, however, remains still open.  In this paper we
``merge'' those two results and show the following:
\begin{Theorem}
  \label{T1n}
  Let $(M,g,F)$ be a solution of the Einstein--Maxwell equations
  containing a connected space-like hypersurface $\Sigma$, the closure
  $\bar \Sigma$ of which is the union of a finite number of
  asymptotically flat ends and of a compact interior.  Let $X$ be a
  Killing vector field on $M$ which is timelike, future directed in
  all the asymptotically flat ends, which leaves $F$ invariant and
  which satisfies the \hso condition \eq{eq:st}. Suppose moreover that:
\begin{enumerate}
 \item We have $g_{\mu\nu}X^\mu X^\nu < 0 $ on\footnote{We use the
    signature $(-,+,+,+)$.} $\Sigma$.
\item The topological boundary $\partial \Sigma\equiv
  \overline{\Sigma}\setminus \Sigma$ of $\Sigma$ is a nonempty
  topological manifold, with $g_{\mu\nu}X^\mu X^\nu = 0 $ on
  $\partial\Sigma$.
  \end{enumerate}
Then:
\begin{enumerate}
\item If $\partial \Sigma$ is \emph{connected}, then $\Sigma$ is
  diffeomorphic to $\R^3$ minus a ball.
  Moreover there exists a neighborhood of $\Sigma$ in $M$ which is
  isometrically diffeomorphic to an open subset of the (extreme or
  non--extreme) \RN space--time.
\item If $\partial \Sigma$ is \emph{not} connected and if condition
  \eq{eq:qc} holds for charges $Q_i$ associated to those components of
  $\partial \Sigma$ that intersect \emph{the degenerate horizons},
  then $\Sigma$ is diffeomorphic to $\R^3$ minus a finite union of
  disjoint balls.  Moreover the space--time contains \emph{only}
  degenerate horizons, and there exists a neighborhood of $\Sigma$ in
  $M$ which is isometrically diffeomorphic to an open subset of the
  \sMP space--time.
\end{enumerate}
\end{Theorem}

Actually a somewhat more general result is proved in Theorem
\ref{T1n1} below. We emphasize that no sign conditions are made
concerning the charges of non--degenerate horizons. We also note that
simple connectedness of $\Sigma$ will hold when appropriate further
global hypotheses on $M$ are done, \emph{cf.} Theorem \ref{T1.1n0}
below. Thus, to obtain a satisfactory classification of the
space--times under consideration it remains to remove the condition on
the sign of the charges, or to construct (and classify) appropriately
regular black holes which do not satisfy this condition. We find that
last possibility rather unlikely.

The definitions and conventions used here coincide with those of the
accompanying paper~\cite{Chstatic}. Those definitions which cannot
be found there are presented in Section \ref{SPnew} below.

We refer the reader to a discussion of a similar theorem for vacuum
space--times in~\cite[Section 1]{Chstatic} for comments concerning
the improvements of this result as compared to the ones available in
the literature even in cases where a mixture of degenerate and
non--degenerate horizons is forbidden. It might be of some interest
to mention that our conclusion will still hold for quite a larger
class of manifolds $\Sigma$. A possible generalization is that with
$\Sigma$ being \emph{e.g.\/} the union of a) a finite number of
asymptotically flat ends with b) a neighborhood of the boundary
$\partial \Sigma$ which has compact closure in $M$ and c) a
non--compact region on which we have $0<\epsilon \le1+ \phi\pm
\sqrt{-g_{\mu\nu}X^\mu X^\nu} $, provided that $\Sigma$ with the
induced metric is a \emph{complete} Riemannian manifold; $\phi$ here
is the electric potential as defined in Equation \eq{eq:pp} after
the relevant duality rotations have been performed, \emph{cf.\/}
Lemma \ref{lnob}. The proof carries through without any
modifications to this case.

Our strategy is a modification of that of Ruback\footnote{We note that
  while the relevant claims in~\cite{Ruback} can be eventually
  justified, the paper~\cite{Ruback} contains several essential gaps.
  The work here can be considered as an extension of that of Ruback to
  include degenerate black holes, together with a justification of the
  relevant unsubstantiated claims made in~\cite{Ruback}.  We further
  note that we have not been able to adapt the technique of Simon
~\cite{Simon:elvac} and Masood--ul--Alam~\cite{Masood} to include
  degenerate black holes without having to introduce some
  supplementary restrictions.}~\cite{Ruback} along the lines of
\cite{Chstatic}: we consider the \emph{orbit space} metric $h$ on
$\Sigma$, as defined in~\cite{Chstatic}.  The key tool here are the
results of~\cite{Chstatic} concerning the geometry of $(\Sigma,h)$
near both the degenerate components of $\partial\Sigma$ and the
non--degenerate ones. Next, following~\cite{Ruback}, we consider a
manifold which consists of two copies of $(\Sigma,h)$ glued along
all \emph{non--degenerate} components of $\partial \Sigma$, equipped
with an appropriate conformally deformed metric. As in
\cite{Chstatic} we use a new version of the positive energy theorem
proved in~\cite{BartnikChrusciel} (Theorem \ref{tpetcharged} below)
to show that the metric on $\Sigma$ is conformally flat. One can
then use classical calculations to finish the proof. We note that it
is usual in the last step of the proof to invoke analyticity to
conclude. Because analytic extensions of manifolds are not unique
this is not sufficient without a more thorough justification. We
finish the proof by a simple open--closed argument which avoids this
problem.

Under the hypotheses of Theorem \ref{T1n} there is no chance of
getting more information about the size of the set on which the
metric is that of a \RN or a \sMP space--time (consider any
hypersurface $\Sigma$ in the \RN space--time, and set $M$ to be any
neighborhood of $\Sigma$ which does not coincide with the \RN
space--time; alternatively, identify $t$ with $t+1$ in the \RN
space--time).  In complete analogy with the vacuum case
in~\cite{Chstatic} we have the following:

\begin{Corollary}\label{c1n}
  Under the hypotheses of Theorem \ref{T1n}, assume further that
  \begin{enumerate}
  \item[3.]  The orbits of the Killing vector $X$ through $\Sigma$ are
    complete.
  \end{enumerate}
  Then the following properties are equivalent:
  \begin{enumerate}
  \item[\phantom{ii}i.] $\Sext$ is achronal$ \!$ {}\footnote{By that
      we mean that there are no timelike curves from $\Sext$ to itself
      which are entirely contained in $\Mext$.} in $\Mext$.
  \item[\phantom{i}ii.] $\Mext$ is diffeomorphic to $\R\times\Sext$
    (which is equivalent to an appropriately complete $\scri$ having
    $\R\times S^2$ topology).
  \item[iii.] There are no closed timelike curves through $\Sext$
    contained in $\Mext$.
  \end{enumerate}
  Further, if one (and hence all) of the above conditions holds, then
  the Killing development$\!$ \footnote{The notion of Killing
    development used here differs slightly from the definition given
    in~\cite{ChBeig1}, as we allow here a topology of $\ks$ which is
    not $\R\times\Sigma$.} $\ks$ of $\Sigma$ defined as
\begin{equation}
\label{kddef}
\ks\equiv\cup_{t\in \R} \phi_t(\Sigma)\ ,
\end{equation}
where $\phi_t$ is the action of the isometry group generated by $X$,
equipped with the induced metric, is isometrically diffeomorphic to a
domain of outer communications of a standard extension of a \RN
space--time or of a \sMP space--time.
\end{Corollary}

The \emph{\sMP} space--times are defined in Section \ref{SPnew}. We
refer the reader to the introduction of~\cite{Chstatic} for a
discussion of the relationship between Theorem \ref{T1n} and
Corollary \ref{c1n} and black holes. In particular in the
introduction of~\cite{Chstatic} an example was given which shows
that more hypotheses than those of Corollary \ref{c1n} are needed to
show that $\ks$ coincides with a d.o.c. in $M$. For reference we
state the following:
\begin{Theorem}
  \label{T1.1n0} Let $(M,g,F)$ be a  solution of the  Einstein--Maxwell
  equations containing a connected space-like hypersurface $\Sigma$,
  the closure $\bar \Sigma$ of which is the union of a finite number
  of asymptotically flat ends and of a compact interior.  Let $X$ be a
  Killing vector field on $M$ which is timelike future directed in all
  the asymptotically flat ends, and which satisfies the \hso
  condition \eq{eq:st}.  Let further $\doc\equiv \doc(\Mext)$ be a
  domain of outer communications in $(M,g)$ associated with one of the
  asymptotically flat ends of $\Sigma$. Suppose that:
\begin{enumerate}
\item We have $\Sigma\subset \doc$.
\item The topological boundary $\partial \Sigma\equiv
  \overline{\Sigma}\setminus \Sigma$ of $\Sigma$ is a nonempty
  topological manifold and satisfies $\partial \Sigma = \overline
  \Sigma \cap \pdoc$.
\item   $X$ has  complete orbits in $\doc$.
\end{enumerate}
\renewcommand{\theenumi}{\alph{enumi}}
\renewcommand{\labelenumi}{4\theenumi)} In addition to the above,
suppose that condition \eq{eq:qc} holds for charges $Q_i$ associated
to those components of $\partial \Sigma$ that intersect \emph{the
  degenerate horizons} and that one of the following conditions holds:
\begin{enumerate}
\item \label{first} 
\label{docgh} Either  $(\doc,g|_{\doc})$ is globally hyperbolic, or
\item \label{Mgh} $(M,g)$ is globally hyperbolic.
\label{last} 
\end{enumerate}
Then the conclusions of Theorem \ref{T1n} and Corollary \ref{c1n}
hold.  Moreover $\doc$ is isometrically diffeomorphic to a domain of
outer communications of a standard extension of a \RN space--time or
of a \sMP space--time.
\end{Theorem}

We note that it is not assumed above that $X$ is timelike throughout
$\Sigma$.

The proofs of both Corollary \ref{c1n} and Theorem \ref{T1.1n0} are
essentially identical to the corresponding ones in~\cite{Chstatic};
some comments about the proof of Corollary \ref{c1n} can be found at
the end of Section \ref{ptt1n}; the proof of  Theorem \ref{T1.1n0}
will be omitted. We note that the property that $\Sigma$ is simply
connected and has only one asymptotically flat end required in
Theorem \ref{T1n} follows from~\cite{ChWald}.  We further note that
the obvious electro--vacuum generalization of the remaining cases of
Theorem 1.3 of~\cite{Chstatic} holds under the supplementary
hypothesis that $\Sigma$ is simply connected and has only one
asymptotically flat end.

This paper is organized as follows: Section \ref{SPnew} contains
definitions and some preliminary remarks. In section \ref{ptt1n} we
prove Theorem \ref{T1n}, as a consequence of the somewhat more general
Theorem \ref{T1n1}, which is also proved there.  \medskip

\noindent\textbf{Acknowledgments:} The author acknowledges useful
discussions with W.~Simon.

\section{Preliminaries}
\label{SPnew} Our conventions and definitions are as in
\cite[Section 2]{Chstatic}. Further, a triple $(M,g,F)$ will be said
to be static if there exists on $M$ a Killing vector field $X$ such
that the Maxwell two--form field $F$ satisfies
$$
{\cal L}_X F = 0\ ,
$$
with $X$ satisfying moreover the \hso condition:
\begin{equation}
  \label{eq:st}
  X_{[\alpha}\nabla_\beta X_{\gamma]} = 0 \ .
\end{equation}
Here and throughout ${\cal L}_X$ denotes the Lie derivative with
respect to $X$.

Next, a data set $(\Sigma _{\mathrm{ext}},g,K)$ with Maxwell field $F$
will be called an \emph{asymptotically flat end} if $\Sigma
_{\mathrm{ext}}$ is diffeomorphic to ${\Bbb R}^3$ minus a ball and if
the fields $(g_{ij},K_{ij})$ satisfy the fall--off conditions
\begin{equation}
|g_{ij}-\delta _{ij}|+r|\partial _\ell g_{ij}|+\cdots +r^k|\partial _{\ell
_1\cdots \ell _k}g_{ij}|+r|K_{ij}|+\cdots +r^k|\partial _{\ell _1\cdots \ell
_{k-1}}K_{ij}|\le C_{k,\alpha}r^{-\alpha }\ ,  \label{falloff}
\end{equation}
for some constants $C_{k,\alpha}$, $\alpha >0$, $k\ge 1$.  We shall
further require that in the local coordinates as above on $\Sext$ the
Maxwell field satisfies the fall--off conditions
\begin{equation}
|F_{\mu\nu}|
+r|\partial _\ell F_{\mu\nu}|+\cdots +r^k|\partial _{\ell
_1\cdots \ell _k}F_{\mu\nu}|
\le \hat C_{k,\alpha}r^{-\alpha-1 }\ ,  \label{Ffalloff}
\end{equation}
for some constants $\hat C_{k,\alpha}$, $\alpha >0$, $k\ge 0$. We
shall always implicitly assume $\alpha >1/2$ when the ADM mass will
be invoked, as this condition makes it well defined in vacuum. It
follows in any case from~\cite[Section 1.3]{Chnohair} that in
stationary electro--vacuum space--times there is no loss of
generality in assuming $\alpha= 1$, $k$ -- arbitrary. A hypersurface
will be said to be \emph{asymptotically flat} if it contains an
asymptotically flat end $\Sigma _{\mathrm{ext}}$.

To avoid ambiguities, we define the \emph{\RN} space--time
$(M^{\mathrm{RN}},g^{\mathrm{RN}})$ to be the manifold $\{t\in\R,r\in
(m+\sqrt{m^2-Q^2-P^2},\infty), q\in S^2\}$, with
\begin{equation}
  \label{eq:chin}
  m^2-Q^2-P^2\ge 0\ ,
\end{equation}
and with the metric
\begin{equation}
  \label{eq:schw}
  g^{\mathrm{RN}}=- (1-\frac{2m}{r}+ \frac{Q^2+P^2}{r^2})dt^2 +
(1-\frac{2m}{r}+ \frac{Q^2+P^2}{r^2})^{-1}dr^2+r^2d\Omega^2\ ,
\end{equation}
where $d\Omega^2$ is the standard round metric on a unit
two--dimensional sphere $S^2$. It is somewhat awkward to build in the
inequality \eq{eq:chin} in our definition of a \RN space--time, but it
saves us the need of repeating that \eq{eq:chin} holds each time we
mention a \RN space--time.  The Maxwell field is
\begin{equation}
  \label{eq:max}
  F\equiv F_{\mu\nu }dx^\mu\wedge dx^\nu = \frac{Q}{r^2}dt \wedge dr
-P \sin(\theta) d\theta \wedge d\phi\ ,
\end{equation}
so that $Q$ is the total electric charge and $P$ is the total magnetic
charge of $\Sext$. We will refer to those coordinates as the standard
coordinates on the \RN space--time. We shall call the \emph{standard
  extension of the \RN space--time} the extension of
$(M^{\mathrm{RN}},g^{\mathrm{RN}})$ described \emph{e.g.\/} by the
Carter--Penrose diagram on page 158 of~\cite{HE} for $m^2>Q^2-P^2$
and on page 160 of~\cite{HE} for $m^2=Q^2-P^2$.

Recall that the Majumdar--Papapetrou (MP) metrics are, locally, of
the form~\cite{Majumdar,Papapetrou:mp} \beqa\label{I.0} & g =
-u^{-2}dt^2
+ u^2(dx^2+dy^2+dz^2)\,, & \\
\label{I.0.1}
&A = u^{-1} dt\,, & \eeqa where $A$ is the Maxwell potential, $F=dA$,
with some nowhere vanishing, say positive, function
$u$. 
A space--time will be called a standard MP space--time if the
coordinates $x^\mu$ of \eq{I.0}--\eq{I.0.1} are global with range
$\R\times(\R^3\setminus\{\vec a_i\})$ for a finite set of points $\vec
a_i\in\R^3$, $i=1,\ldots,I$, and if the function $u$ has the form \beq
\label{standard}
u=1+\sum_{i=1}^I \frac{m_i}{|\vec x - \vec a_i|} \,, \eeq for some
positive constants $m_i$. It has been shown by Hartle and Hawking
\cite{HartleHawking} that every standard MP space--time can be
analytically extended to an electro--vacuum space--time with a
non--empty black hole region, and with a domain of outer communication
which is non--singular in the sense of the theorems proved here. Those extensions will be called the {\em standard} extensions of the \sMP space--times.

\section{Proof of Theorem \protect\ref{T1n}}
\label{ptt1n} Following~\cite{Chstatic}, we equip $\Sigma$ with the
orbit space metric $h$ defined as
\begin{equation}
h(Y,Z)=g(Y,Z)- \frac{g(X,Y)g(X,Z)}{g(X,X)}\ ,
 \label{eq:hdefnew}
\end{equation}
where $X$ is the defining Killing vector, that is, the Killing
vector which asymptotes $\partial/\partial t$ in the asymptotic
regions, and satisfies the \hso condition \eq{eq:st}.  Let the
electric field $E$ and the magnetic field $B$ be defined on $M$ by
the equations (we use the conventions of~\cite{Heuslerbook})
\begin{equation}
  \label{eq:eb}
  E(Y)=-F(X,Y)\ , \qquad B(Y) = (\mystar  F)(X,Y)\ ,
\end{equation}
where $\mystar F$ denotes the space--time Hodge dual of the Maxwell
field two--form $F$.  Simple connectedness of $\Sigma$ and a
standard calculation (\emph{cf., e.g.},~\cite{Heuslerbook}) show
that there exist functions $\phi$ and $\psi$ defined in a
neighborhood of $\Sigma$ in $M$ such that we have
\begin{equation}
  \label{eq:pp}
  E=d\phi\ , \quad B= d\psi\ , \qquad {\cal L}_X\phi = {\cal L}_X\psi=
  0\ .
\end{equation}
By an abuse of notation we shall often use the symbol $\phi$ to denote
the restriction of $\phi$ to $\Sigma$, similarly with $\psi$.  The
potentials $\phi$ and $\psi$ are of course defined up to a constant,
and we can normalize them so that on $\Sext$ we have
\begin{equation}
  \label{eq:ppnorm}
  \phi=\frac{Q}{r}+O(r^{-2})\ , \qquad \psi=\frac{P}{r}+O(r^{-2})\ ,
\end{equation}
where $Q$ is the total electric charge and $P$ is the total magnetic
charge in $\Sext$. (We note if there were several asymptotically flat
ends it could happen that the potentials could asymptote constants
different from zero on some ends, and the proof given below would
break down. This is the only place where the hypothesis that $\Sigma$
has only one end enters in the argument. In fact, one could allow
several ends when the supplementary hypothesis is made that
$\phi$ and $\psi$ can be normalized to asymptote to zero in all
asymptotically flat ends.)

The metric $h$ on $\Sigma$ is essentially ``the metric that would
have been induced on $\Sigma$ if $\Sigma$ were normal to $X$'', so
that we have the following equivalent of Lemma 5.1
of~\cite{Chstatic}, the proof of which is a repetition of that in
\cite{Chstatic}:

\begin{Lemma}
  \label{lfeq} Suppose that $(M,g,F)$ is static and assume that the
  set $(\hat h, \hat V, \hat \phi, \hat \psi)$, where $\hat h$ is the
  metric induced on the hypersurfaces orthogonal to $X$, $-\hat V^2$
  is the square of the Lorentzian norm of $X$ on those hypersurfaces,
  and $ \hat\phi$ (respectively $\hat \psi$) is the restriction of the
  electric potential $\phi$ (respectively the magnetic potential $
  \psi$) defined by Equations \eq{eq:eb} and \eq{eq:pp} to those
  hypersurfaces, satisfies some coordinate--independent system of
  equations. Then the orbit space--metric $h$ together with the
  function $V$ (such that $- V^2$ is the square of the Lorentzian norm
  of $X$ on $\Sigma$), the electric potential $\phi|_{\Sigma}$ and the
  magnetic potential $\psi|_{\Sigma}$ satisfy the same system of
  equations.
\end{Lemma}

It follows that in the Einstein--Maxwell case we have the equations:
\begin{eqnarray}
  \label{eq:fieq}
&   V\Dh \phi = h(d\phi, dV)\ , &
\\ &   V\Dh \psi = h(d\psi, dV)\ , &\label{eq:psieq}
\\ &   V\Dh V = h(d\phi,d\phi )+ h(d\psi,d\psi )\ , & \label{eq:Veq}
\\ &   V
R_{ij}
= D_{i}D_{j}V + V^{-1}\{(h(d\phi,
d\phi)+h(d\psi, d\psi))h_{ij} -2 \phi_{,i}\phi_{,j}-2
\psi_{,i}\psi_{,j}
\}
\ , & \label{eq:Riccieq}
\end{eqnarray}
where $\Dh$ is the Laplace operator of the metric $h$, $R_{ij}$ is the
Ricci tensor of $h$, and where a comma denotes differentiation. In
particular we have
\begin{equation}
  R\equiv h^{ij}R_{ij}= 2V^{-2}\{ h(d\phi,d\phi )+ h(d\psi,d\psi )\}
\label{eq:Ricciscalar}\ .
\end{equation}
Following Heusler~\cite{HeuslerRNuniqueness} we note:
\begin{Lemma}
  \label{lnob} Under the hypotheses of Theorem \ref{T1n}
  the magnetic field $B$ can be made to vanish by a duality rotation.
\end{Lemma}

\proof If $E\equiv0$ the result is obvious by exchanging $\phi$ with
$\psi$. Suppose thus that $E$ is not identically vanishing, as shown
\emph{e.g.\/} in~\cite{HeuslerRNuniqueness} we then have
$$
B  = \mu E \ ,
$$ with $\mu$ being constant on each connected component of the set
$\Omega\equiv\{E\ne 0\}$.  Let $\Omega_0$ be any connected component
of $\Omega$, by performing a duality rotation we can obtain $\psi=0$
in $\Omega_0$~\cite{HeuslerRNuniqueness}.  As $\Omega_0$ is open,
Equation \eq{eq:psieq} and the unique continuation theorem of
Aronszajn~\cite{aronszajn} show that $\psi\equiv0$, hence $B\equiv
0$. \qed

Unless explicitly stated otherwise, in the remainder of the paper we
shall assume that the duality transformation of Lemma \ref{lnob} has
been performed, so that
$$ \psi\equiv 0\ . $$

In the proof of Theorem \ref{T1n} we shall need the following
version of the positive energy theorem, proved
in~\cite{BartnikChrusciel}:

\begin{Theorem}  \label{tpetcharged}
  Let $(\hSigma,\hh)$ be a smooth complete Riemannian manifold with an
  asymptotically flat end $\hSigma_\ext$ (in the sense of Equation
  \eq{falloff} with $k\ge 4$ and $\alpha>1/2$) and with a smooth
  divergence free vector field $\hE$ satisfying
$$
    \hE_idx^i = \frac{\hat Q}{r^2} dr + o(r^{-2})
    $$ in $\hSigma_\ext$. Suppose that the Ricci scalar $\hR$ of $\hh$
    satisfies
  $$ 0 \le \hR- 2\hh(\hE,\hE) \in L^1(\hSigma_\ext) \ .
  $$ Then the ADM mass $\hm$ of $\hSigma_\ext$ satisfies
  $$\hm\ge |\hQ|\ ,
  $$ where $\hQ$ is the total charge of $\hSigma_\ext$.  If the
  equality is attained and $\hat E$ is not identically vanishing, then
  the metric $\hh$ is, locally, the metric induced on the $t=$const
  slices of a \MP space--time (\emph{cf.\/} Equation \eq{I.0}) with
  \begin{equation}
    \label{emp}
    \hE_idx^i = \frac{du}{u} \ ,
  \end{equation}
  where $u$ is as in \eq{I.0}--\eq{I.0.1}.
\end{Theorem}

We emphasize that in the result above $\hSigma$ can have an arbitrary
number (perhaps infinite) of asymptotic ends, and that \emph{no
  hypotheses} are made on the asymptotic behavior of the metric in
those ends except that the metric $\hh$ is complete (and that at
least one of the ends is asymptotically flat so that its ADM mass is
well defined). More general results, allowing for non--vanishing
extrinsic curvature of the initial data hypersurface, non--vanishing
of the magnetic field, poor differentiability of the metric, and
boundaries, can be found in~\cite{BartnikChrusciel}.  The proof of
Theorem \ref{tpetcharged} uses a Witten--type spinorial argument
based on the suggestion of Gibbons and Hull~\cite{GibbonsHull}. The
equality case is handled by the results of Tod~\cite{Tod}; the plane
waves case allowed by Tod is excluded by~\cite[Theorem
3.4]{ChBeig1}. We note that in the $\hat m = |\hat Q|$ case it is
not known whether one can conclude that the metric must be (locally
or globally) a \sMP metric.

To proceed further, we need to analyze the behavior of $h$ and
$\phi$ near $\partial \Sigma$. We shall give here an overview of the
results needed, and we refer the reader to~\cite{Chstatic} for
detailed proofs of the results discussed in this paragraph. Recall,
thus, that by the Vishveshwara--Carter Lemma
\cite{Vishveshwara,CarterJMP} $\partial \Sigma$ must be a subset of
(the closure of) a Killing horizon $\overline{\cal N}$. By that same
lemma one knows that in a static space--time the Killing horizon is
a smooth submanifold.  Standard results~\cite{Heuslerbook} show that
$\phi$ is constant on any connected component of $\overline{\cal
N}$, hence of $\partial \Sigma$. A connected component $S$ of
$\partial\Sigma$ will be called degenerate, respectively
non--degenerate, if $S$ intersects a degenerate, respectively
non--degenerate Killing horizon. By deforming $\Sigma$ slightly in
space--time if necessary we can ensure that $\partial\Sigma$ is a
smooth submanifold both of $\bar \Sigma$ and of $M$ near degenerate
horizons. Every degenerate component corresponds to a complete end
of $(\Sigma,h)$~\cite[Prop.~3.2]{Chstatic}. As far as
non--degenerate horizons are concerned, $\partial \Sigma$ will not
be a smooth submanifold of $M$ in general when there are points on
$\partial \Sigma$ at which the Killing vector field $X$ vanishes.
However we can equip $\bar\Sigma$ with a differentiable structure so
that $\partial \Sigma$ is a smooth submanifold of $\bar \Sigma$
\cite[Prop.~3.3]{Chstatic}.  Moreover $\partial \Sigma$ with this
differentiable structure is a totally geodesic boundary of
$(\Sigma,h)$ across which $h$ can be extended smoothly when doubling
$\Sigma$.  Now $\phi$ is a smooth function on space--time, and the
proof of~\cite[Prop.~3.3]{Chstatic} shows that $\phi$ is a smooth
function of $(x^2,y^a)$ (here $x^2$ denotes the square of $x$, and
not an index $2$ on $x$) in an appropriate coordinate system near a
non--degenerate connected component $S$ of $\partial \Sigma$, with
$S$ given by $x=0$ in this coordinate system. This implies in
particular that
\begin{equation}
  \label{eq:fgrad}
  |d\phi|_h(x=0)=0 \ ,
\end{equation}
and that $\phi$ extends smoothly across $S$ when a doubling of
$\Sigma$ across $S$ is performed.

We have the following, which is based on an observation of {Ruback}
\cite{Ruback}:
\begin{Proposition}
  \label{Pmax}
  Under the hypotheses of Theorem \ref{T1n} we have
\begin{equation}
  \label{eq:**}
  0 \le V + |\phi| \le 1 
\end{equation}
on $\bar \Sigma$, with the inequalities being strict on $\Sigma$
except if the metric is, locally, a Majumdar--Papapetrou metric.
Further the right inequality is strict on non--degenerate horizons.
\end{Proposition}
\proof Set
$$ F_{\pm}= V^2-(1\pm \phi)^2\ ;
$$ as noted by Ruback~\cite{Ruback} the functions $F_{\pm}$ satisfy
the equation
\begin{equation}
  \label{eq:rub}
  \Delta_\gamma F_{\pm}= 0 \ ,
\end{equation}
where $\Delta_\gamma$ is the Laplace operator of the metric
$V^{-2}h_{ij}$. In the asymptotically flat region of $\Sext$ the
$F_{\pm}$'s approach zero, while at every component of
$\partial\Sigma$ we have $F_{\pm} \le 0$.

Suppose, first, that $F_- = 0$ on all components of $\partial \Sigma$;
the maximum principle implies then
$$F_- \equiv 0$$ on $\bar \Sigma$. Equation \eq{eq:Riccieq} and the
transformation rule of the Ricci tensor under conformal
transformations show that the metric $(1-\phi)^{-2}h_{ij}$ is Ricci
flat. In dimension three this implies flatness, and the proof of
Lemma 5.1 of~\cite{Chstatic} shows that near $\Sigma$ the
space--time metric can locally be written in the
Majumdar--Papapetrou form \eq{I.0}. A similar analysis applies if
$F_+$ vanishes throughout $\partial \Sigma$.

It remains to consider the case in which both $F_{+} $ and $F_{-} $
are negative somewhere on $\partial\Sigma$. From the maximum principle
one obtains
\begin{equation}
  \label{eq:Fpm}
  F_{\pm} < 0
\end{equation}
on $\Sigma$, so that
\begin{equation}
  \label{eq:*1}
  V^2< (1-\phi)^2\ , \qquad V^2< (1+\phi)^2
\end{equation}
on $\Sigma$. $V$ has no zeros on $\Sigma$ by hypothesis, which
together with \eq{eq:*1} shows that both $1-\phi$ and $1+\phi$ have no
zeros on $\Sigma$.  As both $1-\phi$ and $1+\phi$ go to $1$ at the
infinity of $\Sext$ it follows that
\begin{equation}
  \label{eq:*3}
  -1 < \phi < 1
\end{equation}
on $\Sigma$. Equations \eq{eq:*1}--\eq{eq:*3} imply $0<V<\min (1+\phi,
1-\phi) = 1-|\phi|$ on $\Sigma$, as desired.

It remains to consider what happens on non--degenerate components of
$\partial \Sigma$. Let, thus, $S$ be a connected non--degenerate
component of $\partial \Sigma$, so that $d\phi$ vanishes on $S$ by
Equation \eq{eq:fgrad}. It is well known, and in any case easily
checked from the formulae in~\cite[Section 3]{Chstatic}, that
\begin{equation}
  \label{eq:keq}
  |dV|_h(x=0)=\kappa \ ,
\end{equation}
where $\kappa$ is the surface gravity of $S$; the condition that $S$
is non--degenerate is precisely $\kappa\ne 0$.  Suppose that $\phi=1$
on $S$, then $F_-$ vanishes on $S$ and Equation \eq{eq:keq} shows that
$F_-=\kappa^2 x^2 + O(x^4)$ will be positive in a neighborhood of $S$
(recall that $\phi-\phi|_{S}=O(x^2)$), which contradicts \eq{eq:Fpm}.
Similarly $\phi=-1$ on $S$ would lead to $F_+$ being positive in a
neighborhood of $S$, again a contradiction.  \phantom{this is to fill
  space}\qed

We note the following corollary\footnote{The inequality \eq{eq:mqrel}
  has been established under rather more general circumstances in~\cite[Remark, p. 107]{Herzlich:mass}, using a technique suggested
by
  Gibbons and Hull~\cite{GibbonsHull}.} of Proposition \ref{Pmax}:

\begin{Corollary}
  \label{c1} Under the hypotheses of Theorem \ref{T1n} we have
  \begin{equation}
    \label{eq:mqrel}
    m \ge |Q|\ .
  \end{equation}
  where $m>0$ is the ADM mass of $\Sigma_{\ext}$ and $Q$ the total
  charge of $\Sext$.  Further, if the inequality is attained the
  metric is, locally, a Majumdar--Papapetrou metric.
\end{Corollary}

\proof A theorem of Beig~\cite{BeigKomar} (\emph{cf.\/} also
\cite{Chremark,ashtekar:magnon:conserved}) shows that the Komar mass
of a static asymptotically vacuum end $(\Sext, g|_{\Sext})$
coincides with its ADM mass, so that we have
 \begin{equation}
   \label{komar}
   V=1-\frac{m}{r} + O(r^{-2})\ .
 \end{equation}
 The inequality \eq{eq:mqrel} follows immediately from Proposition
 \ref{Pmax} and the asymptotic expansion \eq{eq:ppnorm}. If $m=Q$ we
 have $F_+=O(r^{-2})$, and $F_+\equiv 0$ follows from \eq{eq:rub} and
 the asymptotic strong maximum principle of~\cite[Appendix]{Sudarsky:wald}. The conclusion that the metric is
 locally a \MP metric follows then as in the proof of Proposition
 \ref{Pmax}. The case $m=-Q$ follows similarly by considering $F_-$.
 The inequality $m>0$ follows either from the asymptotic strong
 maximum principle of~\cite[Appendix]{Sudarsky:wald} or from
~\cite{Herzlich:mass}.  \qed

 It follows from Proposition \ref{Pmax} that $\phi$ satisfies the
 inequality $ -1 < \phi < 1 $ on $\Sigma$, and that the values
 $\phi=1$ or $\phi=-1$ can only be attained at degenerate components
 of $\partial \Sigma$. When only one component of the event horizon is
 degenerate we can without loss of generality assume, changing $\phi$
 to $-\phi$ if necessary, that we have
\begin{equation}
  \label{eq:goodcase}
  -1 < \phi \le 1 \ \mbox{on} \ \bar\Sigma\ .
\end{equation}
It is tempting to conjecture that one can \emph{always} assume,
changing $\phi$ to $-\phi$ if necessary, that
$$ 0 \le \phi \le 1 \ \mbox{on} \ \bar\Sigma\ .
$$ This is due to the fact that a change of the sign of $\phi$ will
necessarily lead to both positive and negative charges of event
horizons, \emph{cf.\/} Lemma \ref{lopposite} below --- such a
configuration is unlikely to be static. Whatever the situation, if
Equation \eq{eq:goodcase} holds we can prove the following:

\begin{Theorem}
  \label{T1n1}
  Let $(M,g,F)$ be a static solution of the Einstein--Maxwell
  equations with defining Killing vector $X$. Suppose that $M$
  contains a connected and simply connected space--like hypersurface
  $\Sigma$ the closure $\bar \Sigma$ of which is the union of an
  asymptotically flat end and of a compact interior, such that:
\begin{enumerate}
\item We have $g_{\mu\nu}X^\mu X^\nu < 0 $ on $\Sigma$.
\item The topological boundary $\partial \Sigma\equiv
  \overline{\Sigma}\setminus \Sigma$ of $\Sigma$ is a nonempty
  topological manifold, with $g_{\mu\nu}X^\mu X^\nu = 0 $ on
  $\partial\Sigma$.
  \end{enumerate}
  If Equation \eq{eq:goodcase} holds, then the conclusions of Theorem
  \ref{T1n} hold.
\end{Theorem}

\proof The case $m=|Q|=0$ cannot occur by~\cite{Herzlich:mass}. If
$m=|Q|\ne 0$ the metric is, locally, of Majumdar--Papapetrou form by
Corollary \ref{c1}. In that case we can apply~\cite[Theorem
7.2]{ChAsconaa} (\emph{cf.\/} also~\cite{ChNad}) to the Killing
development $(\hat M,\hat g)$ of $\Sigma$ as defined in
\cite{ChBeig1} to conclude that $(\hat M,\hat g)$ is a standard
Majumdar--Papapetrou space--time, and the result follows;
\emph{cf.\/} the argument around Equation \eq{eq:embed} below for a
more detailed exposition of the construction of the embedding in the
\RN context.

It remains to analyze the case $m>|Q|$. In order to do that,
consider the manifold $\Sigma$ equipped with the metric $h$ defined
by Equation \eq{eq:hdefnew}. From what has been said $(\Sigma,h)$ is
a complete Riemannian manifold with compact (perhaps empty) boundary
and with at least one asymptotically flat end $\Sext$.  Let us
denote by $\Snd$ the collection of all those components of the
boundary of $\Sigma$ which correspond to non--degenerate components
of the event horizon of the black hole.  Following~\cite{Ruback}, if
$\Snd\ne\emptyset$ we set
 \begin{eqnarray}
   &\nonumber \Sigma_+ = \Sigma, \qquad h_+ =
   \Big(\frac{1+V+\phi}{2}\Big)^2 h\ , & \\ &\nonumber \Sigma_- =
   \Sigma, \qquad h_- = \Big(\frac{1-V+\phi }{2}\Big)^2 h\ , & \\
   &\label{hh} \hS = 
\Sigma_+ \cup \Sigma_-\cup \Snd
\ ,   \qquad \hat
h\Big|_{\Sigma_+}=h_+\ , \quad 
\hat h\Big|_{\Sigma_-}=h_-\ , & \\ &\label{ehh} \hat E_{\pm}=
\mbox{$\displaystyle\frac{(1+\phi)d\phi - V dV}{V(1+\phi
        \pm V)}$}\ ,
\qquad \hat E\Big|_{\Sigma_+}=\hat E_+\ , \quad 
\hat E\Big|_{\Sigma_-}=\hat E_-\ .&
 \end{eqnarray}
 The topological and differentiable structure of $\hS$ are defined
 through the gluing of $\overline\Sigma_+\equiv\Sigma_+\cup \Snd $
 with $\overline\Sigma_-\equiv\Sigma_-\cup \Snd $ by identifying
 $\Snd$, considered as a subset of $\overline\Sigma_+$, with a second
 copy of $\Snd$, considered as a subset of $\overline\Sigma_-$, using
 the identity map. From our remarks at the beginning of this section
 it follows that the metric $\hat h$ defined on $\Sigma_+\cup\Sigma_-$
 in \eq{hh} can be extended by continuity to a smooth metric on $\hS$;
 similarly $\hat E$ can be extended by continuity to a smooth vector
 field on $\hS$.

 If $\Snd=\emptyset$ we set
 $$\hS=\Sigma\ ,\qquad \hat h=h_+\ ,\qquad \hat E = E_+\ .$$

 We have the following:
 \begin{itemize}
 \item The conformal factor $1- V+\phi\ge 1-V-|\phi|$ is strictly
   positive on $\Sigma\cup \Snd$ by Proposition \ref{Pmax}, and so is
   $1+ V+\phi= 2V+1- V+\phi \ge1- V+\phi$.  Near every connected
   degenerate component $S$ of $\partial \Sigma$ the electric
   potential $\phi$ will tend to a value different from $-1$ by the
   hypothesis \eq{eq:goodcase}, while $V$ will tend to zero, hence the
   asymptotic end of $(\Sigma,h)$ corresponding to $S$ remains
   complete in the metric $(\hat\Sigma,\hat h)$.
 \item The conformal factor $(1+V+\phi)/2$ tends to $1$ in the
   asymptotically flat end $\Sext$, so that $\Sext$ is an
   asymptotically flat end for the metric $h_+$, with ADM mass equal
   to
   $$ \hat m = \frac{1}{2}(m+Q)\ .
   $$ The electric field $\hat E$ approaches zero as $1/r^2$ in
   $\Sext$ and has charge equal to
   $$ \hat Q = \frac{1}{2}(m+Q) = \hat m\ .$$
 \item The conformal factor $1-V+\phi$ tends to $0$ in the
   asymptotically flat $\Sext$ as $(m+Q)/r$, with $m+Q\ne 0$, thus as
   $r$ tends to infinity in $\Sext$ the metric $h_-$ approaches, to
   leading significant orders, the metric
   $$\frac{(m+Q)^2}{4}\Big(\frac{1}{r^2}dr^2+d\Omega^2\Big)\ ,
   $$ where $d\Omega^2$ is the standard round metric on a two sphere.
   It easily follows that $(\Sext,h_-)$ is a \emph{complete} end of
   $(\hat\Sigma,\hat h)$ (\emph{cf.\/} the calculation in the proof of
   Proposition 3.2 in~\cite{Chstatic}).
 \end{itemize}
 As emphasized by Ruback~\cite{Ruback} we have
  \begin{eqnarray}
    \label{eq:hatRe}
    & \hat R = 2 \hat h(\hat E,\hat E)\ , & \\ & \hat \nabla_i \hat
    E^i = 0 \ , \label{eq:hatfe}
  \end{eqnarray}
  where $\hat \nabla$ is the covariant derivative of the metric $\hat
  h$. Thus the hypotheses of Theorem \ref{tpetcharged} are satisfied.
  Since the mass of $\hat h$ and the charge of $\hat E$ coincide,
  Theorem \ref{tpetcharged} shows that $\hat h$ is, locally, the space
  part of the Majumdar--Papapetrou\footnote{It might be worthwhile to
    point out that it is not known at this stage that $\hat h$ is the
    space part of a \sMP metric, but this information is not needed in
    the argument.} metric. This shows in particular that $\hat h$, and
  hence also $h$, are conformally flat, so that the Cotton tensor
  $B_{ijk}$ of $h$ satisfies
  \begin{equation}
    \label{bach}
    B_{ijk}\equiv 0\ .
  \end{equation}
  Equation \eq{emp} implies $d\hE =0$ and from Equation \eq{ehh} we
  have
  \begin{equation}
    \label{emp1}
    0 = d \hE = 2\frac{d\phi\wedge dV}{V^2} \ .
  \end{equation}
  It follows that $d\phi$ is parallel to $dV$ wherever $dV$ does not
  vanish. Standard results about solutions of elliptic equations show
  that
  \begin{equation}
    \label{meq}
    dV = -\frac{mdr}{r^2}+O(r^{-3})\ ,
  \end{equation}
  so that $dV$ does not vanish for $r\ge R$, for an $R$ large enough.
  Increasing $R$ if necessary it follows from Equation \eq{komar} and
  from the maximum principle that for $r\ge R$ the level sets of $V$
  will be embedded spheres. One also finds that there exists $0\le
  V_{-}< 1$ such that for $c\in [V_{-},1)$ the level sets $\{V=c\}$
  are smooth embedded spheres. Let
\begin{eqnarray*}
  \hat I & = & \{c\ | \ c \mbox{ is a non--critical value of } V\}\ ,
\\ \hat\cU & = & \{p \ | \ V(p) \in \hat I\} = \cup_{c\in \hat I}V^{-1}(c)\ ,
\end{eqnarray*}
and define $\cU$ to be that connected component of $\hat \cU$ that
contains $\R^3\setminus B(0,R)$. (Recall that $c$ is non--critical if
$dV$ is nowhere vanishing on the level set $V=c$.) Similarly define
$I\subset (0,1)$ to be that connected component of $\hat
I\setminus\{0\}$ that contains $(V_{-},1)$; clearly
$$
\cU  =  \{p \ | \ V(p) \in I\} = \cup_{c\in I}V^{-1}(c)\ .
$$ Compactness of the level sets of $V$ implies that $\cU$ is
diffeomorphic to $I\times S^2$, and that on $\Uomega $ the function
$V$ can be used as a coordinate. Further we can introduce a finite
number of coordinate patches with coordinates $x^A$, $A=1,2$, on $S^2$
so that on $\Uomega $ the metric takes the form
\begin{equation}
  \label{gcor}
  h=W^{-2} dV^2 + \gamma_{AB}dx^A dx^B \ .
\end{equation}
Equation \eq{emp1} shows that
$$ \phi=\phi(V) $$ on $\Uomega $. This allows one to write Equations
\eq{eq:fieq} and \eq{eq:Veq} in the coordinate system \eq{gcor} as
\begin{eqnarray}
  \label{g1}
  & \mbox{$\displaystyle \frac{\partial \phi}{\partial V}
    = \frac{V}{W \sqrt{\det \gamma_{AB}}} \frac{\partial }{\partial
      V}\Big(W \sqrt{\det \gamma_{AB}} \frac{\partial \phi}{\partial
    V}\Big)$}\ , & \\ & \label{g2}
  \mbox{$\displaystyle \Big(\frac{\partial \phi}{\partial V}\Big)^2 =
    \frac{V}{W
      \sqrt{\det \gamma_{AB}}} \frac{\partial }{\partial V}\Big(W
    \sqrt{\det \gamma_{AB}} \Big)$} \
  . &
\end{eqnarray}
It follows that
$$ V \frac{\partial^2 \phi}{\partial V^2}=\frac{\partial
  \phi}{\partial V} - \Big(\frac{\partial \phi}{\partial V}\Big)^3\ .\
$$ Integrating this equation one finds
\begin{equation}
  \label{g3}
\frac{\partial \phi}{\partial V} = \frac{AV}{\sqrt{1+A^2V^2}}\ ,
\end{equation}
where $A$ is an integration constant. From Equation \eq{meq}, from
$m>0$ and from $d\phi=-Qdr/r^2 +O(r^{-3})$ one obtains
$$ \lim_{V\to 1} \frac{\partial \phi}{\partial V} = -\frac{Q}{m}\ ,
$$ which determines $A$.

Suppose first that $Q=0$, then
$$ d\phi\equiv 0
$$ on $\Uomega $ by Equation \eq{g3}. Equation \eq{eq:fieq} and the
unique continuation theorem of Aronszajn~\cite{aronszajn} show that
$\phi$ is constant throughout $\Sigma$, so that the initial data set
is vacuum. In this case the space--time metric is the Schwarzschild
metric in a neighborhood of $\Sigma$ by~\cite[Theorem
1.1]{Chstatic}.

It remains to consider the case $Q\ne 0$. Integrating Equation \eq{g3}
and using $\lim_{V\to 1} \phi = 0$ we obtain
\begin{equation}
  \label{g5}
  \phi = \frac{m-\sqrt{m^2+Q^2(V^2-1)}}{Q} \ .
\end{equation}
According to Heusler~\cite[Equation (9.58)]{Heuslerbook}
(\emph{cf.\/} also~\cite{MRS:elvac}) this implies
$$
\frac{V^4}{8W^4} B_{ijk}B^{ijk} =
\Big(\frac{m^2-Q^2}{m^2+Q^2(V^2-1)}\Big)^2 \Big(|\lambda|_{\gamma}^2+
\frac{|{\cal D}W|^2_{\gamma}}{2}  \Big)  \ .
$$ Here $|\cdot|_{\gamma}$ denotes the norm with respect to the metric
$\gamma= \gamma_{AB}dx^A dx^B$, ${\cal D}W$ is the gradient of the
restriction of the function $W$ (defined in \eq{gcor}) to the level
sets of $V$, and $\lambda\equiv \lambda _{AB}dx^A dx^B$ is the trace
free part of the extrinsic
curvature tensor of the level sets of $V$ --- in the coordinate system
of \eq{gcor}
\begin{equation}
  \label{lam}
  \lambda _{AB} = W\Big( \frac{\partial \gamma_{AB}}{\partial V}
  -\frac{1}{2}\gamma^{CD}\frac{\partial \gamma_{CD}}{\partial
    V}\gamma_{AB}\Big) \ .
\end{equation}
Equation \eq{bach} implies that $\frac{\partial \gamma_{AB}}{\partial
  V}$ is pure trace, and that $W=W(V)$. This latter property and
Equation \eq{g2} show that $\det \gamma_{AB}$ is a product of a
function of $V$ with a function of the remaining coordinates. From the
asymptotic behavior of the metric it then follows that
\begin{equation}
  \label{gcor2}
  h=W(V)^{-2} dV^2 + H(V)d\Omega^2\ .
\end{equation}
for some function $H(V)$, where $d\Omega^2$ is the standard round
metric on $S^2$. A straightforward integration of Equations \eq{g2}
and \eq{eq:Ricciscalar} using \eq{g5} shows that the metric on
$\Uomega $ is the space part of the Reissner--Nordstr\"om metric. In
other words, $h$ is on $\Uomega $ the pull back by a suitable
diffeomorphism $\psi$ of the space part $h^{\mbox{\tiny RN}}$ of the
Reissner--Nordstr\"om metric.

To finish the proof\footnote{We note that it is usual at this stage to
  invoke analyticity to conclude the proof. Because analytic
  extensions of manifolds are not unique this is not sufficient
  without further justification. The argument we present here avoids
  this problem.}, we claim that $I$ is open in $(0,1)$, which can be
seen as follows: Let $p\in \cU$, we thus have $dV(q)\ne 0$ for all $q$
such that $V(p)=V(q)$. By Equation \eq{gcor2} $|dV|_{h}=W$ is constant
on the level set $V^{-1}(V(p))$ of $V$ through $p$ so that
$$
\inf_{V^{-1}(V(p))}|dV|_{h} 
> 0 \ ,
$$ which easily implies that all nearby level sets are non--critical.

To see that $I$ is closed in $(0,1)$, recall that, using obvious
notation, we have $h=\psi^*h^{\mbox{\tiny RN}} $ and $V=V^{\mbox{\tiny
    RN}}\circ \psi$ on $\cU$.  Let $s_i\in I$ be any sequence
converging to $s\in (0,1)$, thus $s_i=V(p_i)$ for some $p_i\in \cU$.
By the interior compactness of $\Sigma$, passing to a subsequence if
necessary, there exists $p\in\Sigma$ such that $p_i\to p$, with
$V(p)=s>0$. Set
$$ C =\inf |dV^{\mbox{\tiny RN}}|_{h^{\mbox{\tiny RN}}}\ , $$ where
the infimum is taken over those points $q$ in $M^{\mbox{\tiny RN}}$
for which $V^{\mbox{\tiny RN}}(q)>V(p)/2$. We have $V(p_i)=s_i>
V(p)/2$ for $i$ large enough, so that $V^{\mbox{\tiny RN}}(\psi(p_i))
= V(p_i)> V(p)/2$. It follows that
$$ |dV|_{h}(p) = \lim_{i\to \infty } |dV|_{h}(p_i) = \lim_{i\to \infty
  }|dV^{\mbox{\tiny RN}}|_{h^{\mbox{\tiny RN}}}(\psi(p_i)) > C\ , $$
so that $dV(p)\ne 0$. Now $|dV|_h$ is constant on those level sets of
  $V$ which are in $I$, and by continuity it is also constant on those
  level sets of $V$ which are in $\bar I$, the closure of $I$ in
  $(0,1)$. Hence  $|dV|_h$ is non--vanishing on the level set
  $\{V=s\}$, thus $s\in I$.

  We have thus shown that $I$ is open and closed, and connectedness of
  $\Sigma$ implies $\cU=\Sigma$. Thus the manifold $\R\times\Sigma$
  with the metric $-V^2dt^2+h$ is isometrically diffeomorphic to the
  \RN space--time.

  Consider any neighborhood $\cUc$ of $\Sigma$ diffeomorphic to an
  open interval times $\Sigma$; the set $\cUc$ is simply connected by
  simple--connectedness of $\Sigma$. Let $\alpha$ be the one--form
$$
\alpha= \frac{X_\mu dx^\mu}{X_\nu X^\nu}\ ;
$$ Equation \eq{eq:st} shows that $\alpha$ is closed, and
simple--connectedness of $\cUc$ implies existence of a function
$t\in C^\infty (\cUc)$ such that $\alpha=dt$.  As in the proof of
Lemma 5.1 of~\cite{Chstatic} there exists a function $f:\Sigma\to\R$
such that
\begin{equation}
  \label{eq:embed}
  t=s+f\ ,
\end{equation}
Here $s$ denotes the coordinate along the (perhaps only locally
defined) orbits of the Killing vector field on $\cUc$.  Passing to a
subset of $\cUc$ if necessary we may assume that every orbit of $X$ in
$\cUc$ intersects $\Sigma$ precisely once. We can then extend $f$ to a
function on $\cUc$ by requiring that $X(f)=0$. As the metric
$-V^2dt^2+h$ has already been shown to be the \RN metric,
Equation \eq{eq:embed} provides now the required embedding of $\cUc$
into an open subset of the \RN space--time.  \qed

In order to show that Theorem \ref{T1n} is a
  special case of Theorem \ref{T1n1} we need the following result:

\begin{Lemma}\label{lopposite}
  Let $S_a$, $a=1,2$ be connected components of $\partial \Sigma$ such
  that the horizon potentials $\phi_a=\phi|_{S_a}$ satisfy
  \begin{equation}
    \label{eq:minmax}
   \phi_1=\inf_{\bar\Sigma}\phi < 0 \ , \qquad
  \phi_2=\sup_{\bar\Sigma}\phi > 0 \ .
  \end{equation}
  Then the charges $Q_a$ of the $S_a$'s are non--vanishing and have
  opposite signs.
\end{Lemma}
The result it obtained by standard integration by parts arguments.
However, some care must be taken in our context because the degenerate
components of the boundary $\partial \Sigma$ lie at infinite
$h$--distance, and because $V$ tends to zero there.

\medskip

\proof Recall that the charges of the $S_a$'s can be defined by the
equations
\begin{equation}
  \label{eq:charges}
 Q_a=-\lim_{i\to\infty} \int_{S_{a,i}} V^{-1}\nabla^i \phi \,
dS_i\ ,
\end{equation}
where the ${S_{a,i}}$ are any family of connected smooth hypersurfaces
converging in an appropriate sense to the $S_a$'s as $i$ tends to
infinity. For definiteness:
\begin{itemize}
\item If $S_a$ is degenerate we take the ${S_{a,i}}$'s to be the sets
  $x=1/i$, where $x$ is the  coordinate of the proof of
Proposition 3.2 of~\cite{Chstatic}, and we assume that $x$ has been
rescaled so that its range covers the interval $[0,1]$; we set $$
\Omega_a = \{x<1\}\ .
$$
\item If $S_a$ is non--degenerate we take the ${S_{a,i}}$'s to be the
  sets $w=1/i$, where $w$ is the coordinate of the proof of
  Proposition 3.3 of~\cite{Chstatic}, and we assume that $w$ has been
  rescaled so that its range covers the interval $[0,1]$; we set $$
  \Omega_a = \{w<1\}\ .
$$
\end{itemize}
The integrals at the right--hand--side of Equation \eq{eq:charges} are
$i$ independent by equation \eq{eq:fieq} and the divergence theorem,
$$
\int_{S_{a,i}}V^{-1}\nabla^i \phi \,
dS_i-\int_{S_{a,j}}V^{-1}\nabla^i \phi \, dS_i=  \int_{{\cal
      V}_{i,j}} \nabla_i (V^{-1}\nabla^i \phi) \, d\mu_h = 0\ .
$$
Here ${\cal V}_{i,j}$ is the volume the boundary of which consists of
$S_{a,i}$ and  $S_{a,j}$. Hence the limit in \eq{eq:charges} exists.

Let
$$\phi_-=\inf_{S_{1,1}}\phi \ , \qquad \phi_+=\sup_{S_{2,1}}\phi\ . $$
By Equation \eq{eq:minmax} and the maximum principle we have
$\phi_1<\phi_-$ and $\phi_+<\phi_2$.  Let $c$ be a non--critical value
of $\phi$ satisfying $\phi_1<c<\phi_-$,
then the level set $\phi^{-1}\{c\}\cap \Omega_1$ is a smooth compact
submanifold of $\Omega_1$; recall that the set of non--critical values
of $\phi$ is dense by Sard's theorem (\emph{cf., e.g.,}
\cite{GuilleminPollack}).  Applying the divergence theorem on a set
bounded by $\phi^{-1}\{d\}\cap \Omega_1$ (with a non--critical $d$
satisfying $\phi_1<d< \phi_-$) and by $S_{1,i}$ for an $i$ large
enough we obtain
\begin{equation}
  \label{eq:vanish}
  \int_{\phi^{-1}\{d\}\cap \Omega_1}V^{-1}\nabla^i \phi \, dS_i
 = -Q_1\ .
\end{equation}

Let $c$ and $d$ be any non--critical values of $\phi$ satisfying
$\phi_1<c<d<\phi_-$, thus ${\cal W}_{cd}\equiv\{c\le \phi \le d\}
\cap\Omega_1$ is a smooth compact submanifold of $\Omega_1$ with
boundary $(\phi^{-1}\{c\}\cup \phi^{-1}\{d\})\cap\Omega_1$. By the
maximum principle and the boundary point lemma~\cite[Lemma 3.4]{GT}
we have $$h(\nabla \phi, n)>0$$ on $ \phi^{-1}\{d\})\cap\Omega_1$,
where $n$ is that unit normal to $\phi^{-1}\{d\}\cap\Omega_1$ which
points outwards from ${\cal W}_{cd}$, hence
$$Q_1=- \int_{\phi^{-1}\{d\}\cap\Omega_1}V^{-1}\nabla^i \phi \, dS_i <
0\ .$$ The inequality $Q_2>0$ follows by changing $\phi$ to $-\phi$ in
the argument above.  \qed

We can now pass to the

\medskip\noindent{\sc proof of Theorem \protect\ref{T1n}:} If $\partial
\Sigma$ is connected the hypotheses of Theorem \ref{T1n1} are
obviously satisfied, and the result follows. Suppose, thus, that
$\partial \Sigma$ is not connected. Changing $\phi$ to $-\phi$ if
necessary we will be able to satisfy \eq{eq:goodcase} unless there
exists a connected component $S_1$ of $\partial \Sigma$ such that
$\phi_1=-1$ and a connected component $S_2$ of $\partial \Sigma$ such
that $\phi_2=1$. By Proposition \ref{Pmax} $S_1$ and $S_2$ have to be
degenerate, and by Lemma \ref{lopposite} the charges of $S_1$ and
$S_2$ have opposite signs. This is, however, not allowed by the
hypotheses of Theorem \ref{T1n}, and the result follows by Theorem
\ref{T1n1}. \qed

We finally note that the Reissner--Nordstr\"om case of Corollary
\ref{c1n} is proved by a repetition of the arguments of the proof of
Corollary 1.2 in~\cite{Chstatic}. The \MP case is proved by a
repetition of the arguments of the proof of Corollary 1.2 in
\cite{Chstatic} together with the arguments presented in the first
paragraph of the proof of Theorem \ref{T1n1}.

\bibliographystyle{amsplain}
\bibliography{../references/hip_bib,%
../references/reffile,%
../references/newbiblio,%
../references/bibl,%
../references/howard,%
../references/bartnik,%
../references/myGR,%
../references/newbib,%
../references/Energy,%
../references/netbiblio}

\def\polhk#1{\setbox0=\hbox{#1}{\ooalign{\hidewidth
  \lower1.5ex\hbox{`}\hidewidth\crcr\unhbox0}}}
  \def\polhk#1{\setbox0=\hbox{#1}{\ooalign{\hidewidth
  \lower1.5ex\hbox{`}\hidewidth\crcr\unhbox0}}} \def\cprime{$'$}
  \def\cprime{$'$} \def\cprime{$'$} \def\cprime{$'$}
\providecommand{\bysame}{\leavevmode\hbox to3em{\hrulefill}\thinspace}
\providecommand{\MR}{\relax\ifhmode\unskip\space\fi MR }
\providecommand{\MRhref}[2]{%
  \href{http://www.ams.org/mathscinet-getitem?mr=#1}{#2}
}
\providecommand{\href}[2]{#2}
\begin{thebibliography}{10}

\bibitem{aronszajn}
N.~Aronszajn, \emph{A unique continuation theorem for solutions of elliptic
  partial differential equations or inequalities of second order}, Jour.\ de
  Math. \textbf{XXXVI} (1957), 235--249.

\bibitem{ashtekar:magnon:conserved}
A.~Ashtekar and A.~Magnon-Ashtekar, \emph{On conserved quantities in general
  relativity}, Jour.\ Math.\ Phys. \textbf{20} (1979), 793--800.

\bibitem{BeigKomar}
R.~Beig, \emph{{Arnowitt--Deser--Misner energy and $g_{00}$}}, Phys.\ Lett.
  \textbf{69A} (1978), 153--155.

\bibitem{ChBeig1}
R.~Beig and P.T. Chru\'{s}ciel, \emph{Killing vectors in asymptotically flat
  space-times: {I. A}symptotically translational {K}illing vectors and the
  rigid positive energy theorem}, Jour.\ Math.\ Phys. \textbf{37} (1996),
  1939--1961, arXiv:gr-qc/9510015.

\bibitem{CarterJMP}
B.~Carter, \emph{Killing horizons and orthogonally transitive groups in
  space-time}, Jour.\ Math.\ Phys. \textbf{10} (1969), 70--81.

\bibitem{Chremark}
P.T. Chru\'{s}ciel, \emph{A remark on the positive energy theorem}, Class.\
  Quantum Grav. \textbf{33} (1986), L115--L121.

\bibitem{Chnohair}
\bysame, \emph{``{N}o {H}air'' {T}heorems -- folklore, conjectures, results},
  Differential Geometry and Mathematical Physics (J.~Beem and K.L. Duggal,
  eds.), Cont.\ Math., vol. 170, AMS, Providence, 1994, pp.~23--49,
  arXiv:gr--qc/9402032.

\bibitem{Chstatic}
\bysame, \emph{The classification of static vacuum space-times containing an
  asymptotically flat spacelike hypersurface with compact interior}, Class.\
  Quantum Grav. \textbf{16} (1999), 661--687, \emph{Corrigendum} in
  arXiv:gr-qc/9809088v2.

\bibitem{ChNad}
P.T. Chru\'{s}ciel and N.S. Nadirashvili, \emph{All electrovacuum
  {M}ajumdar--{P}apapetrou space-times with non--singular black holes}, Class.\
  Quantum Grav. \textbf{12} (1995), L17--L23, arXiv:gr-qc/9412044.

\bibitem{ChWald}
P.T. Chru\'{s}ciel and R.M. Wald, \emph{On the topology of stationary black
  holes}, Class.\ Quantum Grav. \textbf{11} (1994), no.~12, L147--152,
  arXiv:gr--qc/9410004. \MR{MR1307013 (95j:83080)}

\bibitem{GibbonsHull}
G.W. Gibbons and C.M. Hull, \emph{A {B}ogomolny bound for general relativity
  and solitons in {$N=2$} supergravity}, Phys.\ Lett. \textbf{109B} (1982),
  190--194.

\bibitem{GT}
D.~Gilbarg and N.S. Trudinger, \emph{Elliptic partial differential equations of
  second order}, Springer, Berlin, 1983.

\bibitem{GuilleminPollack}
V.~Guillemin and A.~Pollack, \emph{Differential topology}, Prentice--Hall,
  Englewood Cliffs, N.J, 1974.

\bibitem{HartleHawking}
J.B. Hartle and S.W. Hawking, \emph{Solutions of the {E}instein--{M}axwell
  equations with many black holes}, Commun.\ Math.\ Phys. \textbf{26} (1972),
  87--101.

\bibitem{HE}
S.W. Hawking and G.F.R. Ellis, \emph{The large scale structure of space-time},
  Cambridge University Press, Cambridge, 1973, Cambridge Monographs on
  Mathematical Physics, No. 1. \MR{MR0424186 (54 \#12154)}

\bibitem{Herzlich:mass}
M.~Herzlich, \emph{The positive mass theorem for black holes revisited}, Jour.\
  Geom.\ Phys. \textbf{26} (1998), 97--111.

\bibitem{HeuslerRNuniqueness}
M.~Heusler, \emph{On the uniqueness of the {R}eissner--{N}ordstr{\"{o}}m
  solution with electric and magnetic charge}, Class.\ Quantum Grav.
  \textbf{11} (1994), L49--L53.

\bibitem{Heuslerbook}
\bysame, \emph{Black hole uniqueness theorems}, Cambridge University Press,
  Cambridge, 1996.

\bibitem{heuslerMP}
\bysame, \emph{On the uniqueness of the {P}apapetrou--{M}ajumdar metric},
  Class.\ Quantum Grav. \textbf{14} (1997), L129--L134, arXiv:gr-qc/9607001.

\bibitem{Israel:elvac}
W.~Israel, \emph{Event horizons in static electrovac space-times}, Commun.\
  Math.\ Phys. \textbf{8} (1968), 245--260.

\bibitem{Majumdar}
S.D. Majumdar, \emph{A class of exact solutions of {E}instein's field
  equations}, Phys. Rev. \textbf{72} (1947), 390--398.

\bibitem{Masood}
A.K.M. Masood{--ul--A}lam, \emph{Uniqueness proof of static charged black holes
  revisited}, Class.\ Quantum Grav. \textbf{9} (1992), L53--L55.

\bibitem{MRS:elvac}
H.~M{\"uller zum Hagen}, D.C. Robinson, and H.J. Seifert, \emph{Black holes in
  static electrovac space-times}, Gen.\ Rel.\ Grav. \textbf{5} (1974), 61--72.

\bibitem{Papapetrou:mp}
A.~Papapetrou, \emph{A static solution of the equations of the gravitational
  field for an arbitrary charge distribution}, Proc. Roy. Irish Acad.
  \textbf{A51} (1947), 191--204.

\bibitem{Ruback}
P.~Ruback, \emph{A new uniqueness theorem for charged black holes}, Class.\
  Quantum Grav. \textbf{5} (1988), L155--L159.

\bibitem{Simon:elvac}
W.~Simon, \emph{Radiative {E}instein-{M}axwell spacetimes and `no--hair'
  theorems}, Class.\ Quantum Grav. \textbf{9} (1992), 241--256.

\bibitem{Sudarsky:wald}
D.~Sudarsky and R.M. Wald, \emph{Extrema of mass, stationarity and staticity,
  and solutions to the {E}instein--{Y}ang--{M}ills equations}, Phys.\ Rev.
  \textbf{D46} (1993), 1453--1474.

\bibitem{Tod}
K.P. Tod, \emph{All metrics admitting super--covariantly constant spinors},
  Phys. Lett. \textbf{121B} (1983), 241--244.

\bibitem{Vishveshwara}
C.V. Vishveshwara, \emph{Generalization of the {``{S}chwarzschild Surface''} to
  arbitrary static and stationary metrics}, Jour.\ Math.\ Phys. \textbf{9}
  (1968), 1319--1322.

\end{thebibliography}

\end{document}